\documentclass[prl,amsmath,amssymb,twocolumn,superscriptaddress]{revtex4}
\usepackage{amsmath, amstext, amssymb, amsfonts, amsxtra}
\usepackage[usenames]{color}
\usepackage{grffile}
\usepackage{bm}
\newcommand{\como}{Center for Nonlinear and Complex Systems, Dipartimento
di Scienza e Alta Tecnologia, Universit\`a degli Studi dell'Insubria,
via Valleggio 11, 22100 Como, Italy}
\newcommand{\infn}{Istituto Nazionale di Fisica Nucleare, Sezione di Milano,
via Celoria 16, 20133 Milano, Italy}
\newcommand{\NEST}{NEST, Istituto Nanoscienze-CNR, I-56126 Pisa, Italy}
\newcommand{\brazil}{International Institute of Physics, Federal University
of Rio Grande do Norte, Campus Universit\'ario - Lagoa Nova, CP. 1613,
Natal, Rio Grande Do Norte 59078-970, Brazil}
\newcommand{\xiamen}{Department of Physics and Fujian Provincial Key
Laboratory of Low Dimensional Condensed Matter Physics, Xiamen University,
Xiamen 361005, Fujian, China}
\newcommand{\lanzhou}{Lanzhou Center for Theoretical Physics, Lanzhou
University, Lanzhou 730000, Gansu, China}

\begin{document}

\title{Rectification of heat current in Corbino geometry}

\author{Zhixing Zou}
\affiliation{\xiamen}
\author{Giulio Casati}
\affiliation{\como}
\affiliation{\brazil}
\author{Giuliano Benenti}
\affiliation{\como}
\affiliation{\infn}
\affiliation{\NEST}
\author{Jiao Wang}
\affiliation{\xiamen}
\affiliation{\lanzhou}

\begin{abstract}
We prove analytically the ballistic thermal rectification effect (BTRE) in the Corbino disk
characterized by an annular shape. We derive the thermal rectification efficiency (RE) and
show that it can be expressed as the product of two independent functions, the first dependent
on the temperatures of the heat baths and the second on the system's geometry. It follows that
a perfect BTRE can be reached with the increase of the ratios of the heat baths' temperatures
and of the radius of the outer edge to the inner edge of the disk. We also show that, by
introducing a potential barrier into the Corbino disk, the RE can be greatly improved. Quite
remarkably, by an appropriate choice of parameters, the thermal diode effect can be reversed.
Our results are robust under variation of the Corbino geometry, which may provide a novel and
flexible route to manipulate the heat flow at the nanoscale.
\end{abstract}

\maketitle

\textit{Introduction.}-- The control and management of the heat current is becoming increasingly
important for the future society. In this context, the possibility of building devices capable
of rectifying the heat current has been demonstrated~\cite{Peyrard2002, Peyrard2006}. Various
mechanisms  have been suggested and investigated in order to increase the rectification
efficiency (RE), by magnifying the spatial dependence of the local thermal conductivity through
inhomogeneity or asymmetry of the material structure (see~\cite{Walker2011, Li2012, BCMP2016,
Wehmeyer2017, BDLL2023} and references therein). In spite of the fact that the laws of physics
do not put limitations to devise efficient thermal rectifiers, we are still far from a
satisfactory understanding of this phenomenon both analytically and experimentally.

The rapid development of nanotechnology has led to the consideration of the ballistic thermal
rectification effect (BTRE)~\cite{Wu2021, Ouyang2010, Zhang2010}, which has attracted a lot of
interest recently, in view of its potential implications for designing novel thermal nanodevices.
Indeed, when the system size is comparable to the phonon mean free path, ballistic transport may
dominate in lattices~\cite{CWZhang2013, Volz2019, Zardo2020}. The basic idea goes back to Song
{\it et al}~\cite{Song1998}, who considered the ballistic transport of electrons in a GaAs-AlGaAs
heterostructure and proposed the ballistic rectifier, which relies on a new kind of rectification
mechanism that is entirely different from the ordinary electrical diode. A photon based thermal
rectifier in which all thermal energy transfer takes place through vacuum has been proposed
in~\cite{Otey2010} as well. In~\cite{Schmotz2011}, instead, a thermal rectification device
based on standard silicon processing technology has been demonstrated, where the heat flow
is carried by ballistic phonons in a thin Si membrane.

The Corbino geometry~\cite{Corbino1911} has been considered to study the quantum Hall
effect~\cite{Dolgopolov1992} and, more recently, the Nernst effect caused by magnetization
currents flowing along the inner and outer edges of the Corbino disk, maintained at different
temperatures~\cite{Altshuler2020}. The thermoelectric response of Corbino structures has also
been measured, and Corbino devices have been envisioned as thermoelectric coolers at low
temperatures~\cite{Arrachea2020}.

In this paper, we study ballistic thermal transport in the Corbino geometry. First, we
analytically calculate the heat currents and show the presence of a strong rectification when
the temperatures of the inner and outer edges of the disk are interchanged. To our knowledge
this is the first case for which the RE is rigorously calculated. Second, we show that the
addition of a potential barrier inside the Corbino disk can strongly enhance the rectification
effect. Finally, quite surprisingly, there are parameter regions where the thermal diode
effect can be reversed by varying either the height or the position of the barrier.

\textit{The model.}-- We consider the billiard model of the Corbino geometry, i.e., point
particles moving freely inside an annular disk (see Fig.~1). The inner and the outer circular
edges are in contact with two thermal baths at temperature $T_i$ and $T_o$, respectively.
When a particle collides with an edge, it is reflected back with a random velocity according
to the distribution~\cite{bath1, bath2}
$\bm \Pi_\alpha (\bm{v})= \Pi_\alpha (v,\theta)=P_\alpha (v) \tilde P(\theta)$, with
$\tilde P(\theta)=\frac{1}{2}\cos\theta$ and
\begin{equation}
P_\alpha (v)=\sqrt{\frac{2m}{\pi k_B T_\alpha}}\frac{m v^2}
{k_B T_\alpha}e^{-\frac{m v^2}{2k_B T_\alpha}}.
\end{equation}
Here, the subscript $\alpha=i$ or $o$, indicates the edge, inner or outer, from which the
particle is reflected back, $v=|\bm{v}|$ denotes the magnitude of the reflection velocity,
$-\frac{\pi} {2} < \theta < \frac{\pi}{2}$ represents the angle between the reflected velocity
$\bm{v}$ and the normal direction at the colliding point, $m$ is the particle mass, and
$k_B$ is the Boltzmann constant.

\begin{figure}
\label{fig:model}
\centering
\includegraphics[width=8cm]{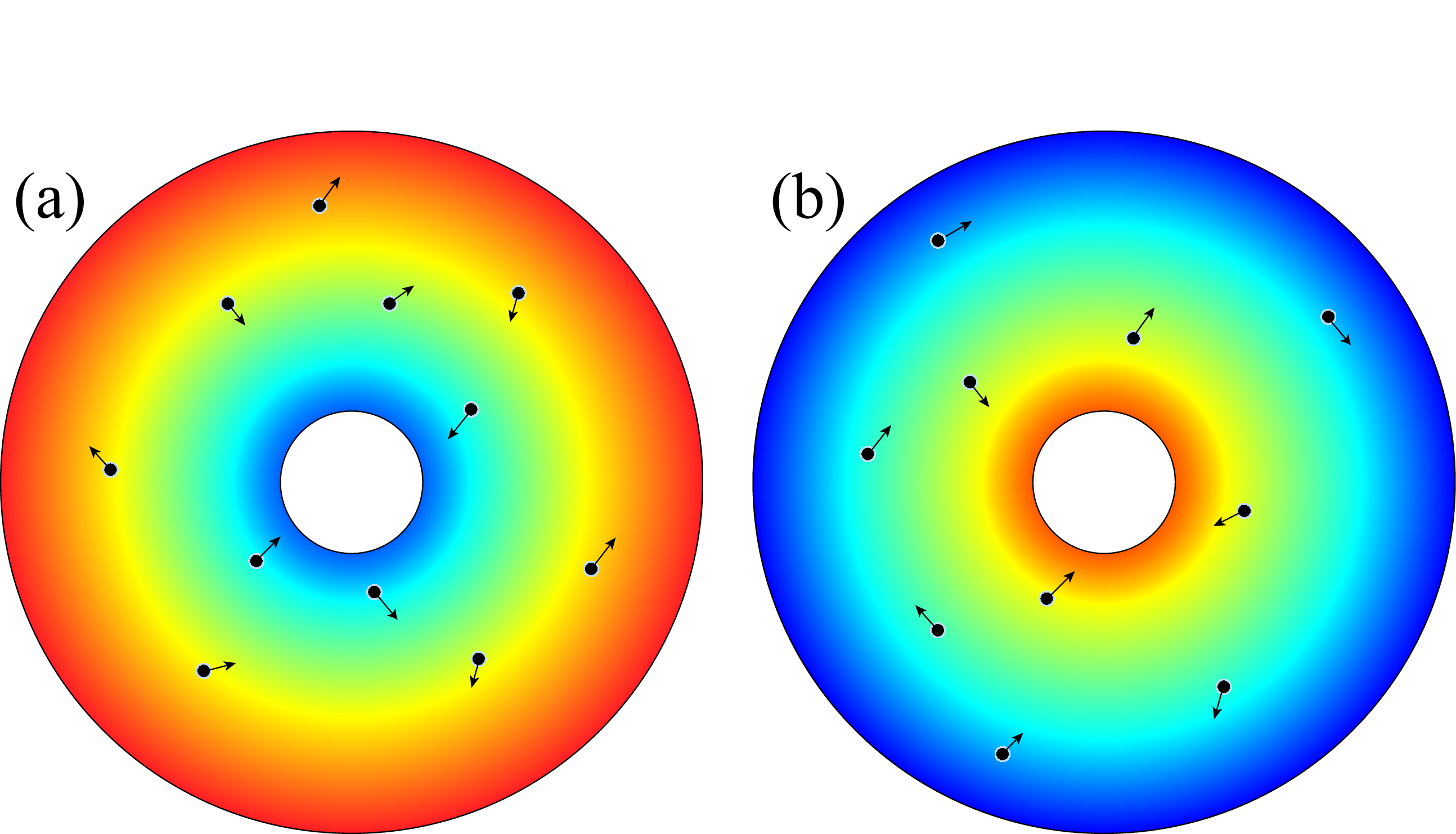}
\caption{Schematic plots of the Corbino billiard model for asymmetric thermal transport.
(a) The setup when the hot (cold) bath is coupled to the outer (inner) edge and a `forward
flux', $J_f$, forms in the stationary state. (b) The setup when the two baths in (a) are
swapped and a `reverse flux',  $J_r$, forms, instead. The color changes from blue to red
as the average kinetic energy of the particles increases.}
\end{figure}

At the stationary state, the heat flux that flows across the system from the hot to the cold
bath can be written as
\begin{equation}\label{eq:flux definition}
J=N\left|\frac{\langle E_{o\to i}\rangle-\langle E_{i\to o}\rangle}
{\langle t_{o\to i}\rangle+\langle t_{i\to o}\rangle}\right |,
\end{equation}
where $N$ is the number of particles in the system, $\langle E_{o\to i}\rangle$ and $\langle
t_{o\to i}\rangle$ represent, respectively, the average energy a particle transfers and the
average time it takes during a journey from the outer to the inner edge. Note that during
such a journey, the particle may collide with the outer edge one or more times. The quantities
$\langle E_{i\to o}\rangle$ and $\langle t_{i\to o}\rangle$ have similar meanings but for the
journey from the inner to the outer edge, instead. As the inner edge is convex, a particle
that leaves from it will reach the outer edge straightforwardly.

Let us denote the temperature of the hot and the cold baths by $T_H$ and $T_C$ ($T_H>T_C$),
respectively. When we set $T_o=T_H$ and $T_i=T_C$, the heat flux from the hot outer bath to
the cold inner bath will be denoted by `forward flux', $J_{f}$. Instead, when $T_i=T_H$ and
$T_o=T_C$, the heat flux from the hot inner bath to the cold outer bath  will be denoted by
`reverse flux', $J_{r}$. In what follows, we will derive an explicit expression for the RE
defined as  $|\xi |$, with
\begin{equation}\label{eq:xi}
  \xi=\frac{J_{f}-J_{r}}{J_{f}+J_{r}}.
\end{equation}
It is clear that $0\le |\xi| \le 1$, and the larger $|\xi|$, the stronger the thermal
rectification effect, with perfect rectification corresponding to $|\xi|=1$. Due to the
geometric asymmetry along the flux direction, we can intuitively anticipate that $J_{r} < J_f$,
so that $\xi>0$. Indeed, in the reverse configuration, the inner edge is hot and more particles
aggregate near the outer edge as their velocities are lower after colliding with the cold and
longer outer edge. Therefore, the particles that per unit time actively transport the energy
are in effect fewer than in the forward configuration.

\textit{Theoretical analysis.}-- The average energy transferred by a particle during a journey
between the two baths can be written as
\begin{equation}\label{eq:eio}
\langle E_{o\to i, i\to o}\rangle = \int_0^\infty \frac{1}{2}mv^2 P_{o,i}(v) dv
=\frac{3}{2}k_B T_{o,i}.
\end{equation}
To compute $\langle t_{o\to i}\rangle$, let us denote by $\varsigma$ a journey path from the
outer to the inner edge, by $\overline{t_{o\to i}(\varsigma)}$  the average time a particle spends
to go through the path $\varsigma$, and by $p_{o\to i}(\varsigma)$ the probability for a particle
to take the path $\varsigma$.
Then $\langle t_{o\to i}\rangle = \sum\overline{t_{o\to i}(\varsigma)}p_{o\to i}(\varsigma)$,
where the summation runs over all the allowed paths. Note that the dependence of $\langle t_{o\to i}
\rangle$ on $v$ appears only in $\overline{t_{o\to i}(\varsigma)}$, while $p_{o\to i}(\varsigma)$
is completely determined by the angle(s) at which the particle leaves from the outer edge every
time after it collides with the latter during the journey. Therefore, if the length of the path
$\varsigma$ is $d_{o\to i}(\varsigma)$, then $\overline{t_{o\to i}(\varsigma)}$ can be expressed
as $\overline{t_{o\to i}(\varsigma)} = \int  [{d_{o\to i}(\varsigma)}/{v}] P_o(v)dv$ such that
\begin{equation}\label{eq:result_tio}
\langle t_{o\to i}\rangle=\langle d_{o\to i}\rangle \sqrt{\frac{2m}{\pi k_B T_o}},
\end{equation}
with $\langle d_{o\to i}\rangle=\sum_{\varsigma} d_{o\to i}(\varsigma) p_{o\to i} (\varsigma)$
being the average length of all the paths from the outer to the inner edge. For the average
time a particle takes to travel from the inner to the outer edge, $\langle t_{i\to o}\rangle$,
we have the similar result. Finally, considering Eq.~\eqref{eq:flux definition}, we have
\begin{equation}\label{eq:form_J}
J=N\left|\frac{\frac{3}{2}k_B\left(T_o-T_i\right)}{\langle d_{o\to i}
\rangle\sqrt{\frac{2m}{\pi k_B T_o}}+\langle d_{i\to o}\rangle
\sqrt{\frac{2m}{\pi k_B T_i}}}\right|
\end{equation}
and, by substituting into Eq.~\eqref{eq:xi} the corresponding $J_f$ and $J_r$ (given by
Eq.~\eqref{eq:form_J} with $T_{o,i}=T_{H,C}$ and $T_{o,i}=T_{C,H}$, respectively),
\begin{equation}\label{eq:result_xi}
\xi=\frac{\sqrt{T_H}-\sqrt{T_C}}{\sqrt{T_H}+\sqrt{T_C}}\cdot\frac{\langle d_{o\to i}\rangle
-\langle d_{i\to o}\rangle}{\langle d_{o\to i}\rangle +\langle d_{i\to o}\rangle}.
\end{equation}
Note that $\xi$ is the product of two independent functions: one is exclusively related to the
temperatures of the two baths and the other is exclusively determined by the geometry of the
model. Obviously, in the limit $\langle d_{o\to i}\rangle/\langle d_{i\to o}\rangle \to \infty$,
$\xi$ reaches its maximum value
\begin{equation}\label{eq:xi_max}
  \xi_{max}=\frac{\sqrt{T_H}-\sqrt{T_C}}{\sqrt{T_H}+\sqrt{T_C}}.
\end{equation}
Furthermore, as $T_H/T_C\to \infty$, $\xi_{max}\to 1$, implying that perfect rectification
is achievable in our model.

Due to the circular symmetry of the annular Corbino geometry, the average length of the
paths between the two edges can be derived explicitly (see Supplemental Material~\cite{SM}),
leading to
\begin{equation}\label{eq:heq0-xi}
\xi=\xi_{max}\cdot \frac{\left(\pi-2\Theta_o \right) r^2_o-2\sqrt{r^2_o-1}}{\pi(r^2_o-1)},
\end{equation}
where $r_o=R_o/R_i>1$, with $R_o$ and $R_i$ being the radius of the outer and the inner
edge, respectively, and $\Theta_o=\arcsin(1/r_o)$. It is straightforward to show that both
$\langle d_{o\to i}\rangle$ and $\langle d_{i\to o} \rangle$ are monotonically increasing
functions of $r_o$, $\langle d_{o\to i}\rangle>\langle d_{i\to o} \rangle$,
and $\langle d_{o\to i} \rangle/\langle d_{i\to o} \rangle \to \infty$ as $r_o\to \infty$.
As a consequence, $\xi$ is also a monotonically increasing function of $r_o$ and as
$r_o\to \infty $, it approaches its maximal value $\xi_{max}$ given by Eq.~\eqref{eq:xi_max}.
The analytical results for $\xi$ are compared with the simulation results in Fig.~2 (see the
blue line and symbols). We can see that they agree with each other perfectly.

\begin{figure}
\includegraphics[width=9.0cm]{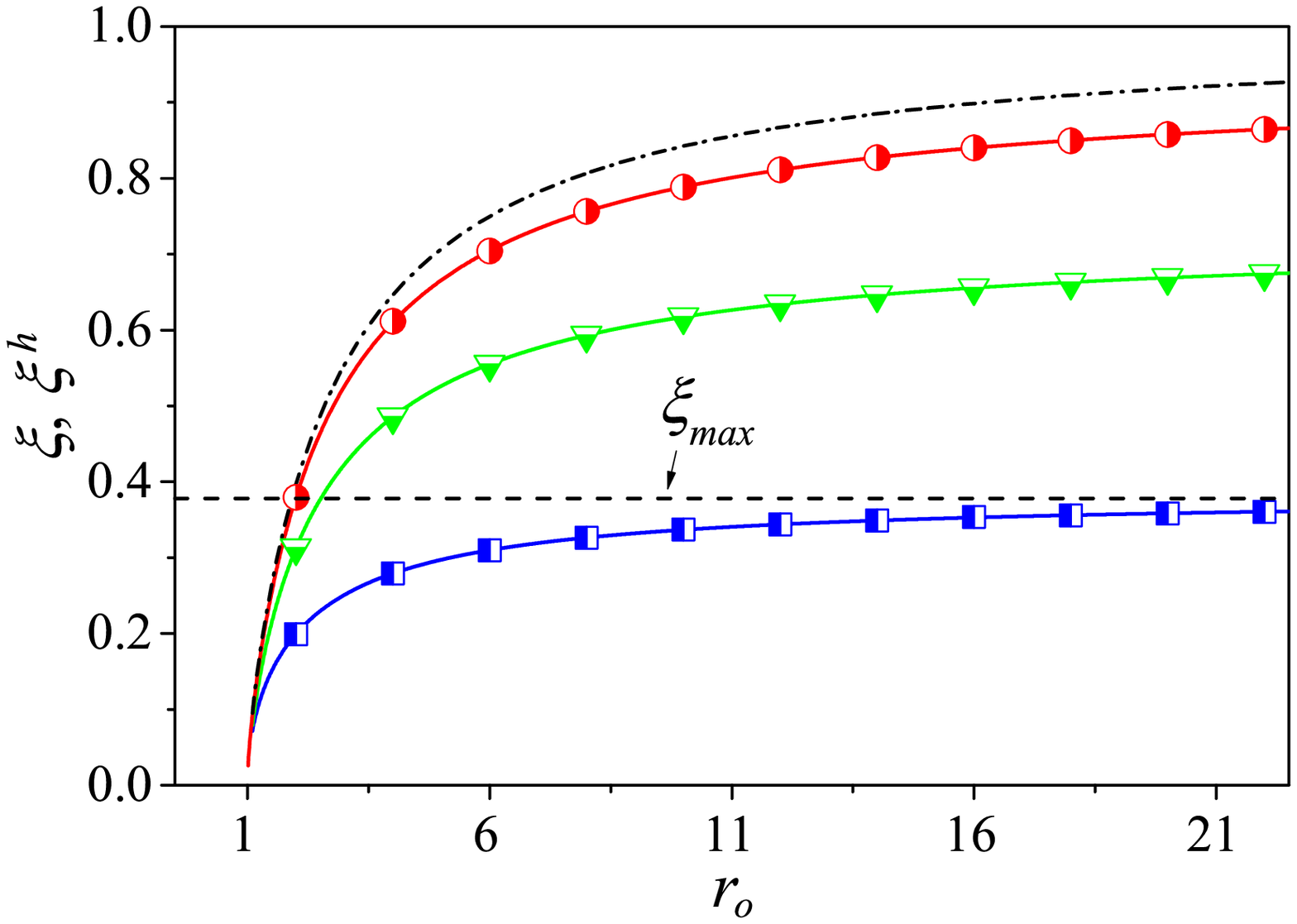}
\caption{The rectification efficiency as a function of $r_0=R_o/R_i$ for the (generalized)
Corbino model. The solid lines are for the analytical results and the accompanying symbols
are for the simulation results. The blue line is for $h=0$, while the green and red lines are
for $h=2$ and $h=4$, respectively. The dashed line indicates the value of $\xi_{max}$ given by
Eq.~\eqref{eq:xi_max} and the dash-dotted line corresponds to the limiting case of $h\to \infty$
given by Eq.~\eqref{eq:xi-infty}. Here, $T_C=1$, $T_H=5$, $R_i=1$, and $R_b=(R_i+R_o)/2$. Note
that for all simulations throughout, the particle mass, the particle number density, and the
Boltzmann constant are fixed to be unity.}
\label{fig:annular}
\end{figure}

Note that, remarkably, the obtained expression for $\xi$ in Eq.~(\ref{eq:result_xi}) has
a general validity beyond the specific annular Corbino disk. For given $T_c$ and $T_h$,
it allows us to search for the geometry with a larger ratio of
$\langle d_{o\to i}\rangle/\langle d_{i\to o}\rangle$ for a better RE. For instance, keeping
the area enclosed by the outer edge unchanged but changing its shape from a circle to an ellipse,
the RE can be improved significantly (see Supplemental Material~\cite{SM}).

\textit{A generalized Corbino model.}-- As mentioned previously, changing the geometry may
improve the RE, but it is upper bounded by $\xi_{max}$. Is it possible to overcome this
bound? Here we show that the answer is positive if an `energy filtering mechanism' is
exploited. This can be illustrated by introducing a potential barrier of width zero and height
$h$ in our model. For convenience of analysis, the potential barrier is assumed to be located
on a concentric circle of radius $R_b$ ($R_i<R_b<R_o$). When a particle hits the barrier, it
will pass through if its kinetic energy in the normal direction at the hitting point is larger
than $h$. Otherwise, it will be reflected back specularly. As only the particles that can
pass the barrier are effective energy carriers, the existence of the barrier provides an
additional tool to modulate the asymmetry between the forward and the reverse heat fluxes.

The original Corbino model is a special case of this generalized one with $h=0$. Thanks to
the circular symmetry of the system, the heat flux and the RE for the general case $h\ne 0$
can be derived as well (see Supplemental Material~\cite{SM}).
But in the expression of the RE, dependence on bath temperatures and geometry parameters are
intertwined with each other, in clear contrast to the case $h=0$ where they play their roles
independently. At any rate, we can derive intriguing analytical results, which are corroborated
by numerical simulations. In the following, we will use $\xi^h$ to denote the RE for a given
value of $h$.

First of all, for a given finite value of $r_b=R_b/R_i$, as $r_o\to \infty$, the saturation
value of the RE, denoted as $\xi_{max}^h$, satisfies $\xi_{max}^h>\xi_{max}$ for any $h>0$
($\xi_{max}$ is given by Eq.~\eqref{eq:xi_max} and corresponds to the case of $h=0$). In
particular,
\begin{equation}
\xi_{max}^{h} = \frac{\sqrt{T_H} P\left(T_H\right)-\sqrt{T_C} P\left(T_C\right)}{\sqrt{T_H}
P\left(T_H\right)+\sqrt{T_C} P\left(T_C\right)},
\end{equation}
where $P(T)= 1 +
r_b e^{-\frac{h}{k_B T}} \mbox{Erf}\left( \sqrt{\frac{h}{k_B T}\frac{1}{r_b^2-1}}\right)-
\mbox{Erf}\left(r_b \sqrt{\frac{h}{k_B T}\frac{1}{r_b^2-1}}\right)$, which is a monotonically
increasing function of $T$. Therefore, $P(T_C)/P(T_H) <1$, and as a result, $\xi_{max}^h>\xi_{max}$
(see Supplemental Material~\cite{SM}).
In Fig.~2, the RE as a function of $r_o$ for various potential heights is presented. It can
be seen that, by introducing the potential barrier, the RE can be greatly improved.

Our main theoretical results are then summarized in Fig.~3(a), where the rectification factor
$\xi^h$ as a function of both the position and the height of the potential barrier is presented.
The $r_b - h$ space can be divided into three regions with distinctive features, separated by
the critical values $r_b=r_b^*$ and $r_b=r_b^\dagger$, respectively, where $r_b^*$ and
$r_b^\dagger$ correspond to the two ends of the black dotted curve defined by
$\partial \xi^h / \partial h=0$. Therefore, in region I ($1<r_b<r_b^*$) and region II
($r_b^*<r_b<r_b^\dagger$) above the dotted line, $\xi$ monotonically increases with $h$,
while in region II below the dotted line and in region III ($r_b^\dagger<r_b<r_o$), $\xi$
monotonically decreases with $h$. The black dashed line is for the curve $\xi^h=0$. It separates
region III in two parts, where in the top half (blue color) reverse rectification ($\xi<0$) occurs,
whose RE increases with $h$. Fig.~3(b) and (c) show the $h$ dependence of the fluxes and the
RE $\xi^h$, respectively. The transition from forward to reverse rectification is observed.
Finally, note that $r_b^\dagger$ can also be determined as the solution of $\xi^\infty=0$,
so that the reversible thermal diode is found in the entire region III.

In the limit of $h\to \infty$, a simple analytical expression for the RE can be derived,
\begin{equation}\label{eq:xi-infty}
\xi^\infty=\frac{\pi(r^2_o-r^2_b)-(2r^2_b\Theta_b-\pi+2\sqrt{r^2_b-1})}
{\pi(r^2_o-r^2_b)+(2r^2_b\Theta_b-\pi+2\sqrt{r^2_b-1})},
\end{equation}
where $\Theta_b=\arcsin(1/r_b)$. Note that $\xi^\infty$ is independent of the bath temperatures.
From Eq.~(\ref{eq:xi-infty}) we can infer that when the position of the potential barrier is close
to the inner or the outer edge of the system, i.e., $r_b\to 1$ or $r_b\to r_o$, the perfect rectification
is approached, in the former case with $J_f > J_r$ and  $\xi^\infty\to 1$, in the latter in the
reversed mode  $J_r > J_f$ and $\xi^\infty\to -1$. With increasing $h$, there is a trade-off between
increasing the RE and decreasing the heat current, as also clear from Fig.~3(b) and (c). For a large
but finite value of $h$, the analytical expression $\xi^\infty$ serves to obtain a good approximation
for the corresponding RE.

\begin{figure}[t!]
\includegraphics[width=9.3cm]{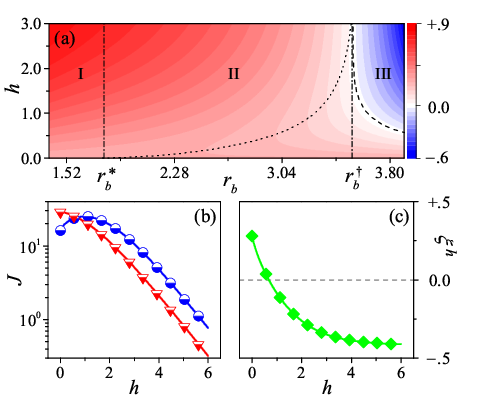}
\caption{ (a) Dependence of the thermal rectification efficiency $\xi^h$ on $h$ and $r_b=R_b/R_i$.
The black dotted line corresponds to $\partial \xi^h/\partial h=0$ and the black dashed line
corresponds to $\xi^h=0$. (b) Dependence of the forward (red) and the reverse (blue) flux on
$h$ for a case in the reversible diode region III with $r_b=3.8$. (c) The thermal rectification
efficiency corresponds to (b). In (b) and (c), the solid lines (symbols) are for the analytical
(numerical) results. Here, $T_C=1$, $T_H=5$, $R_i=1$, and $R_o=4$.}
\end{figure}

\textit{Discussions and Conclusions.}-- In summary, by studying the ballistic thermal
rectification effect in the Corbino disk,
the analytical expressions of the heat flux and the thermal rectification
efficiency have been derived and corroborated by the simulation results.
In particular, we have provided a positive
answer to the question if the perfect thermal rectification can be reached, in principle.

The models we have considered are two dimensional (2D). However, the results can be extended
to three dimension (3D) straightforwardly. For example, for the 3D counterpart of the Corbino
model where the boundaries and the potential barrier are three concentric spheres, analytical
results for the heat flux and the thermal rectification efficiency can also be obtained (see
Supplemental Material~\cite{SM}). In particular, for the case without potential barrier, we
have $\xi=\xi_{max} \cdot g(R_o/R_i)$, with $g(r_o)=(r_o^2-1)^{3/2} /(r_o^3-1)$. Again, $g(r_o)$
is a monotonically increasing function of the radius ratio $r_o=R_o/R_i$ that saturates to one
as $r_o\to \infty$. Note that in this 3D Corbino model, as the asymmetry in terms of $\langle
d_{o\to i} \rangle /\langle d_{i\to o}\rangle$ is much stronger than in the original 2D model
for the same radius ratio $r_o$, the thermal rectification efficiency is much stronger as well.

The obtained analytical results for the 2D Corbino model also apply to the 2D model of a fan
shape, where the two (left and right) thermal baths are coupled to the two arc boundaries,
respectively, with the other two sides being adiabatic. Similarly, the obtained analytical
results for the 3D counterpart of the Corbino model apply to the 3D model of partial spheres
(of the same solid angle) for the two edges and the potential barrier.

Qualitatively, the thermal rectification characteristics revealed by the 2D Corbino model
and its 3D counterpart should be shared by the billiard models of the same topology, i.e.,
with two loop curves (closed surfaces) as an inner and an outer edge coupled with two heat
baths, respectively, and one loop curve (closed surface) in between where a potential barrier
is located. This generality might facilitate experimental studies.

A further non-trivial extensions of our model could be obtained by considering thermochemical
baths, exchanging both heat and particles with the system, thus exploring in Corbino-like
geometries the possibility of a diode that rectifies both particle and heat currents. Finally,
it would be interesting to explore quantum (wave) diodes with the same geometries.

Energy has become a major issue in modern society, and one of its crucial elements is thermal
management. However, thermal engineering has not benefited, so far, from ingenious devices as
electrical diodes. Efficient thermal rectifiers would greatly contribute to a variety of
applications ranging from nanoscale heat regulation, to solar-thermal power devices, heat
engines, refrigerators, up to thermal management of buildings. With the present study we hope
to attract attentions and stimulate experimental work in this direction.

We acknowledge support by the National Natural Science Foundation of China (Grants
No. 12075198, No. 12247106, and No. 12247101), the  Julian Schwinger Foundation
(Grant JSF-21-04-0001), the PRIN MUR (Grant No. 2022XK5CPX), and the INFN through
the project QUANTUM.


\appendix

\begin{widetext}

\begin{center}
{\bf Supplementary material: Rectification of heat current in Corbino geometry}
\end{center}

Here, we provide the detailed derivation of the current and the rectification efficiency (RE) in the original two dimensional (2D) Corbino disk, the generalized 2D Corbino disk with a potential barrier, and in their three dimensional (3D) counterparts. In addition, the RE of the disk model that has a circular inner edge and an elliptical outer edge, is investigated numerically and compared with that of the Corbino disk.

In the following, for a particle departing from the inner edge, we will use $L_{io}(\theta)$ and $L_{ibi}(\theta)$, respectively,  to denote the
distance it travels to reach the outer boundary, and the distance it travels to the barrier, being reflected back, and returns to the inner edge. For a particle departing from the outer edge, we will use $L_{oi}(\theta)$, $L_{obo}(\theta)$ and $L_{oo}(\theta)$, respectively, to denote
the distance it travels to reach the inner edge, the distance it travels to the barrier, being reflected back, and return to the outer edge, and the distance it travels from the outer edge to the outer edge. Based on geometric relations, we have
\begin{equation}
L_{io}(\theta)=\sqrt{R_o^2-R_i^2\sin^2\theta}-R_i\cos\theta,
\end{equation}
\begin{equation}
L_{ibi}(\theta)=2(\sqrt{R_b^2-R_i^2\sin^2\theta}-R_i\cos\theta),
\end{equation}
\begin{equation}
L_{oi}(\theta)=R_o\cos\theta-\sqrt{R_i^2-R_o^2\sin^2\theta},
\end{equation}
\begin{equation}
L_{obo}(\theta)=2(R_o\cos\theta-\sqrt{R_b^2-R_o^2\sin^2\theta}),
\end{equation}
and
\begin{equation}
L_{oo}(\theta)=2R_o\cos\theta.
\end{equation}
Note that these relations hold for both the 2D and the 3D cases.

When the potential barrier of height $h$ and radius $R_b$ is introduced, we will use $v_{ib}(\theta)$ ($v_{ob}(\theta)$) to denote the minimum speed for the particle departing from the inner (outer) edge with angle $\theta$ that can pass the barrier. It follows that
\begin{equation}
v_{ib}(\theta)=\sqrt{\frac{2h R_b^2}{m\left(R_b^2-R_i^2\sin^2\theta\right)}}
\end{equation}
and
\begin{equation}
v_{ob}(\theta)=\sqrt{\frac{2h R_b^2}{m\left(R_b^2-R_o^2\sin^2\theta\right)}}.
\end{equation}

\section{The Corbino disk without potential barrier}
In this case, the average
length $\langle d_{i\to o}\rangle$ of all the paths from the outer to the inner edge can be obtained straightforwardly:
\begin{equation}\label{eq:avetio}
\begin{split}
  \langle d_{i\to o}\rangle &=  \int_{-\frac{\pi}{2}}^{\frac{\pi}{2}}
   L_{io}(\theta)\tilde{P}_i(\theta) \mathrm{d}\theta =\frac{R_i}{2}\left[r_o^2\Theta_o+\sqrt{r_o^2-1}-\dfrac{\pi}{2}\right],
  \end{split}
\end{equation}
where $r_o\equiv R_o/R_i$ and $\Theta_{o}\equiv \arcsin(R_i/R_o)$.
However, for the average length $\langle d_{o\to i}\rangle$ of the paths from the outer to the inner edge
it is not so straightforward, as the particle may collide with the outer edge multiple times before reaching the inner edge.
If we use $p_n$ to denote the probability for a particle to take a journey that starts from the outer edge, collides with the outer boundary $n-1$ times, before reaching the inner edge, and $d_n$ the averaged distance it travels during such a journey, we have $\langle d_{o\to i}\rangle=\sum
_{n=1}^{\infty} p_n d_n$. For a particle starting from the outer edge, suppose the probability it collides with the inner edge or the outer edge next is
$p_{oi}$ or $p_{oo}$, respectively, and the distance it travels during such a free motion is $d_{oi}$ or $d_{oo}$, respectively, then by definition we have
\begin{equation}\label{eq:heq0-poi}
p_{oi} =\int_{-\Theta_{o}}^{\Theta_{o}} \tilde{P}_i(\theta) \mathrm{d}\theta = r_o^{-1},
\end{equation}
\begin{equation}\label{eq:heq0-poo}
p_{oo}=\int_{-\frac{\pi}{2}}^{-\Theta_{o}}
\tilde{P}_o(\theta) \mathrm{d}\theta + \int_{\Theta_{o}}^{\frac{\pi}{2}}
\tilde{P}_o(\theta)\mathrm{d}\theta  =1-r_o^{-1},
\end{equation}
\begin{equation}\label{eq:heq0-toi}
\begin{split}
  d_{oi} &=\frac{1}{p_{oi}}\int_{-\Theta_{o}}^{\Theta_{o}}  L_{oi}(\theta)\tilde{P}_o(\theta) \mathrm{d}\theta\\
  &=\frac{R_i}{2}\left(r_o^2\Theta_{o}+\sqrt{r_o^2-1}-\frac{\pi}{2}\right),
 \end{split}
\end{equation}
and
\begin{equation}\label{eq:heq0-too}
\begin{split}
  d_{oo} &=\frac{1}{p_{oo}}{\left(\int_{-\frac{\pi}{2}}^{-\Theta_{o} } \mathrm{d}\theta  + \int_{\Theta_{o} }^{\frac{\pi}{2}} \mathrm{d}\theta \right)  \left( L_{oo}(\theta)\tilde{P}_o(v,\theta)\right)} \\
  &=\frac{R_i}{2(r_o-1)}\left[{\left(\pi-2\Theta_{o}\right)r_o^2 - 2\sqrt{r_o^2-1}}\right].
\end{split}
\end{equation}

For the sake of clarity, we will use the notation $\int_{oi} \mathrm{d}\theta$ and
$\int_{oo} \mathrm{d}\theta$, respectively, to denote the integration of $\theta$ over the angle range
where the particle will collide with either the inner boundary or the outer boundary. 
Then $p_n d_n$ can be expressed as
\begin{equation}\label{eq:heq0-tn}
\begin{split}
 p_n d_n  &={\int_{oo} \mathrm{d}\theta_1 \cdots\int_{oo} \mathrm{d}\theta_{n-1} \int_{oi} \mathrm{d}\theta_{n}
           \left( \left(\sum\limits_{i=1}^{n-1}L_{oo}\left(\theta\right)+L_{oi}\left(\theta\right)\right)\prod\limits_{i=1}^{n}\tilde{P}_o(\theta_i)\right)}\\
       &=(n-1)\left( \int_{oo} \tilde{P}_o(\theta)\mathrm{d}\theta\right)^{n-2}  \int_{oo}  L_{oo}\left(\theta\right)\tilde{P}_o(\theta)\mathrm{d}\theta \int_{oi} \tilde{P}_o(\theta)\mathrm{d}\theta + \left( \int_{oo}  \tilde{P}_o(\theta)d\theta\right)^{n-1}  \int_{oi}  L_{oi}\left(\theta\right) \tilde{P}_o(\theta)d\theta\\
       &={(n-1)p_{oo}^{n-1}p_{oi}d_{oo} + p_{oo}^{n-1}p_{oi} d_{oi}}.
  \end{split}
\end{equation}
Then $\langle d_{o\to i}\rangle$ can be expressed as
\begin{equation}\label{eq:heq0-avetoi}
\begin{split}
  \langle d_{o\to i}\rangle &= \sum\limits_{n=1}^{\infty} p_n d_n\\
                                    &=\sum\limits_{n=1}^{\infty}(n-1)p_{oo}^{n-1}p_{oi}d_{oo} + \sum\limits_{n=1}^{\infty} p_{oo}^{n-1}p_{oi} d_{oi}\\
                                    &= \frac{p_{oo}}{p_{oi}}d_{oo} + d_{oi}.
\end{split}
\end{equation}

What Eq.~\eqref{eq:heq0-avetoi} suggests is clear: For a particle starting from the outer
edge and reaching the inner edge in the end, it will experience ${p_{oo}}/{p_{oi}}$ times
collisions with the outer edge on average. So the average distance the particle travels can be divided into two parts as the
r.h.s. of Eq.~\eqref{eq:heq0-avetoi} indicates.

Substituting Eq. \eqref{eq:heq0-poi}, \eqref{eq:heq0-poo}, \eqref{eq:heq0-toi}, and
\eqref{eq:heq0-too} into \eqref{eq:heq0-avetoi}, the final result of $\langle d_{o\to i}\rangle$
is
\begin{equation}\label{eq:avetoi}
\langle d_{o\to i}\rangle=\frac{R_i}{2}\left[r_o^2\left( \pi - \Theta_{o}\right)
-\sqrt{r_o^2-1}-\frac{\pi}{2}\right].
\end{equation}
By substituting $\langle d_{i\to o}\rangle$ and $\langle d_{o\to i}\rangle$ into Eq.~(7) of the main text, then the RE can be obtained (Eq.~(9) of the main text).

\section{The disk with elliptical outer edge}
For the case without the potential barrier, the only strategy to increase the RE at given heat baths' temperatures is to increase the ratio $\langle d_{o\to i}\rangle/\langle d_{i\to o}\rangle$ by adjusting the shapes of the edges. Here we present an example, where the outer circular edge of the Corbino disk is replace by an ellipse, as shown in Fig. \ref{fig:ellipse model}. The length of the semi-major and the semi-minor axis is denoted by $a$ and $b$, respectively. When $a=b=R_o$, it reduces to the Corbino disk. For comparison with the latter, we set $ab=R_o^2$, i.e., the area enclosed by the outer edge in the two cases is the same.
\begin{figure}[!h]
  \includegraphics[width=12cm]{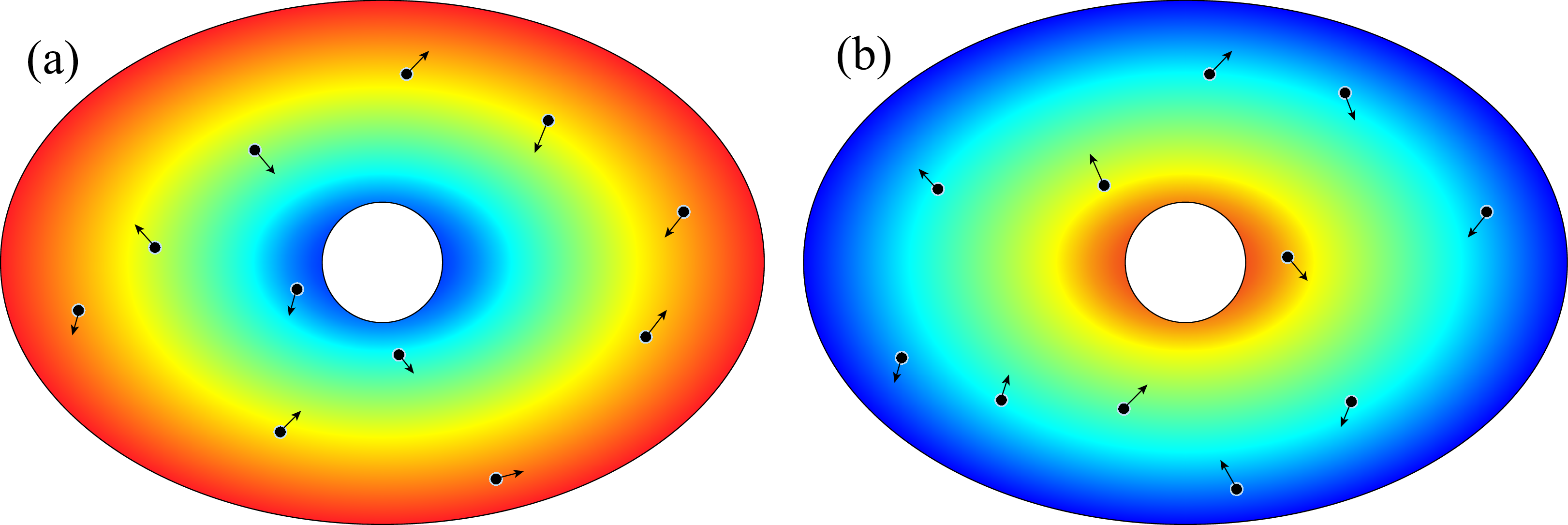}
  \caption{Schematic plots of the disk with an elliptical outer edge. (a) The outer (inner) edge is in contact with the hot (cold) bath, where a forward flux  (from the outer to the inner bath) forms.  (b) The reverse case of (a), where a reverse flux forms. }
\label{fig:ellipse model}
\end{figure}

The results are shown in Fig. \ref{fig:compare}. In panel (a) we fix $ab=R_o^2=25$ and change the ratio $a/b$. Note that the leftmost data point for $a/b=1$ is for the result of the Corbino disk. It can be seen that, as $a/b$ increases,  the RE increases progressively. In panel (b), we fix $b=1.01R_i$ but change $R_o$ (meanwhile $ab=R_o^2$). We can see that the RE of the disk with the elliptical outer edge is always greater than that of the Corbino disk, and as $R_o$ increases, both approach the maximum value $\xi_{max}$.
\begin{figure}[!h]
\hskip-1cm
    \includegraphics[width=12cm]{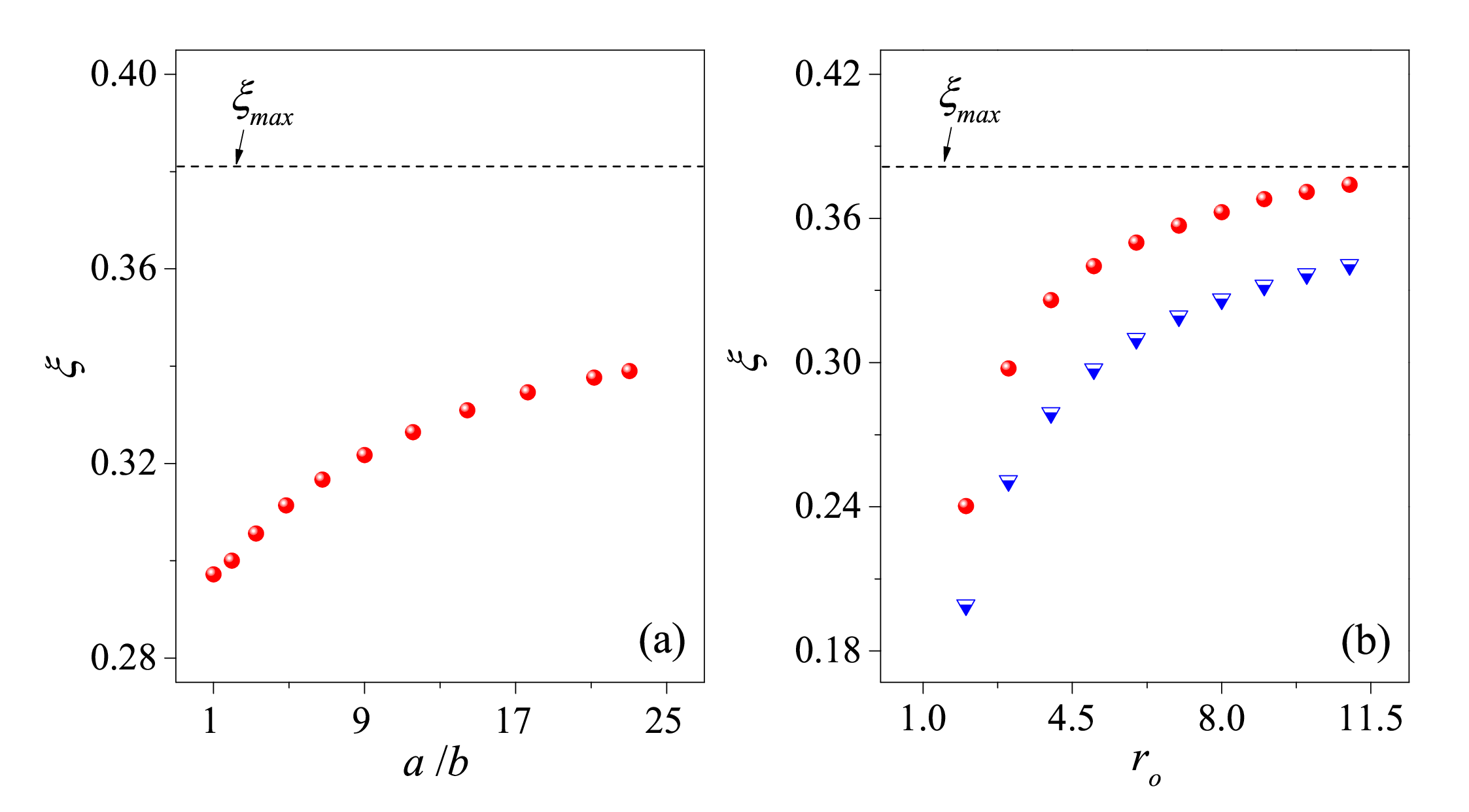}
  \caption{(a) The thermal rectification efficiency of the disk model with an elliptical outer edge with a fixed area $\pi ab=25\pi$. (b) Comparison of the thermal rectification efficiency between the Corbino disk (blue triangles) and the disk with an elliptical outer edge (red dots). Here, $r_o=R_o/R_i$, $b=1.01R_i$, and $a=R_o^2/b$. For all the simulations, $T_H=5$, $T_C=1$, and $R_i=1$.}
\label{fig:compare}
\end{figure}

\section{The 3D counterpart of the Corbino model}
In the 3D counterpart of the Corbino disk, the inner and the outer edges are two concentric spheres of radius $R_i$ and $R_o$, respectively, where two heat baths of temperature $T_i$ and $T_o$ are brought into contact. The velocity a particle takes when it is reflected back at edge $\alpha$ follows the distribution
\begin{equation}
\mathcal{P}_\alpha (v,\theta,\phi) =\frac{m^2}{2\pi k_B^2 T^2} v^3 \sin\theta \cos\theta e^{-\frac{m v^2}{2 k_B T_\alpha}},
\end{equation}
where $\theta$ is the polar angle and $\phi$ is the azimuth angle with respect to the normal direction at the reflecting point. As the distribution is isotropic
in $\phi$, $\mathcal{P}_\alpha (v,\theta,\phi)$ reduces to
 \begin{equation}
\Pi_\alpha^{3D} (v,\theta) =\frac{m^2}{k_B^2 T^2} v^3 \sin\theta \cos\theta e^{-\frac{m v^2}{2 k_B T_\alpha}}.
\end{equation}
Taking the same definitions of $\langle E_{i\to o}\rangle$, $\langle E_{o\to i}\rangle$,
$\langle t_{i\to o}\rangle$, and $\langle t_{o\to i}\rangle$ as in the 2D case,
we have
\begin{equation}\label{eq:eitoo_3d}
\langle E_{i\to o}\rangle =\int_{0}^{\frac{\pi}{2}}\mathrm{d}\theta \int_{0}^{\infty} \frac{1}{2}mv^2 \Pi_i^{3D}(v,\theta) \mathrm{d}v \\
=2k_B T_i,
\end{equation}
\begin{equation}\label{eq:eotoi_3d}
 \langle E_{o\to i}\rangle =\frac{\displaystyle{\int_{0}^{\Theta_{o}} \mathrm{d}\theta \int_{0}^{\infty} \frac{1}{2}mv^2 \Pi_o^{3D}(v,\theta) \mathrm{d}v}}{p_{oi}} \\
  =2k_B T_o,
\end{equation}
\begin{equation}\label{eq:titoo_3d}
\begin{split}
  \langle t_{i\to o}\rangle &=  \int_{0}^{\frac{\pi}{2}}\mathrm{d}\theta \int_0^\infty
  \frac{L_{io}(\theta)}{v}\Pi_i^{3D}(v,\theta) \mathrm{d}v \\
  &=\frac{R_i}{3}\sqrt{\frac{\pi m}{2 k_B T_i}}\left[{r_o^3-\left(r_o^2-1\right)^{3/2}-1}\right],
  \end{split}
\end{equation}
and
\begin{equation}\label{eq:totoi_3d}
\begin{split}
  \langle t_{o\to i}\rangle &=\frac{p_{oo}}{p_{oi}}t_{oo} + t_{oi}\\
  &=\frac{R_i}{3}\sqrt{\frac{\pi m}{2 k_B T_o}}\left[{r_o^3+\left(r_o^2-1\right)^{3/2}-1}\right],
\end{split}
\end{equation}
where $t_{oo}$ and $t_{oi}$ represent the averaged time a particle spends when it starts from the outer edge and collides
with the outer edge and the inner edge, respectively. 
In addition,
\begin{equation}
  p_{oi} =\int_{0}^{\Theta_{o}} \mathrm{d}\theta \int_{0}^{\infty} \Pi_i^{3D}(v,\theta) \mathrm{d}v\\
  =r_o^{-2},
\end{equation}
\begin{equation}
  p_{oo} =\int_{\Theta_{o}}^{\frac{\pi}{2}} \mathrm{d}\theta \int_{0}^{\infty} \Pi_i^{3D}(v,\theta) \mathrm{d}v\\
  =1-r_o^{-2},
\end{equation}
\begin{equation}
\begin{split}
  t_{oi} &=\frac{1}{p_{oi}}{\int_{0}^{\Theta_{o}} \mathrm{d}\theta \int_{0}^{\infty} \frac{L_{oi}(\theta)}{v}\Pi_o^{3D}(v,\theta) \mathrm{d}v}\\
  &=\frac{R_i}{3}\sqrt{\frac{\pi m}{2 k_B T_o}}\left[{r_o^3-\left(r_o^2-1\right)^{3/2}-1}\right],
 \end{split}
\end{equation}
and
\begin{equation}
\begin{split}
  t_{oo} &=\frac{1}{p_{oo}}{\int_{\Theta_{o}}^{\frac{\pi}{2}} \mathrm{d}\theta \int_{0}^{\infty}  \frac{L_{oo}(\theta)}{v}\Pi_o^{3D}(v,\theta) \mathrm{d}v}\\
  &=\frac{2}{3}R_i\sqrt{\frac{\pi m}{2 k_B T_o}}  \left(r_o^2-1\right)^{1/2}.
 \end{split}
\end{equation}
Thus, the final results for $J$ and $\xi$ are, respectively,
\begin{equation}
J= N \sqrt{\dfrac{2 k_B T_i T_o}{\pi m}} \frac{6 k_B\left| T_o-T_i\right|}{R_i\left\lbrace\sqrt{T_o}\left[r_o^3-\left(r_o^2-1\right)^{3/2}-1\right]+\sqrt{T_i}
\left[r_o^3+\left(r_o^2-1\right)^{3/2}-1\right]\right\rbrace}
\end{equation}
and
\begin{equation}
\xi= \frac{\sqrt{T_H}-\sqrt{T_C}}{\sqrt{T_H}+\sqrt{T_C}}\cdot \frac{\left(r_o^2-1\right)^{\frac{3}{2}}}{r_o^3-1}.
\end{equation}

\section{The generalized Corbino disk with a potential barrier}
Suppose that the the potential barrier has a zero width, a height of $h$, and locates at the concentric circle of radius $R_b$ . In this case, for a particle departing from the inner boundary, either it travels to the outer edge directly, or it is reflected back by the barrier and returns to the inner edge. We use $p_{io}$ and $p_{ibi}$ to denote the probability, and $t_{io}$ and $t_{ibi}$ to denote the average time for such a particle to take the two possibilities, respectively. Similarly, for  a particle departing from the outer edge, it may travel to the inner edge directly, or travel to the outer edge directly, or be reflected back by the barrier and return to the outer edge. We use $p_{oi}$, $p_{oo}$, and $p_{obo}$ to denote the probability,  and $t_{oi}$, $t_{oo}$, and $t_{obo}$ to denote the average time, respectively, for one such particle to take the three possibilities.  Then, by definition, $\langle E_{i\to o}\rangle$ and $\langle E_{o\to i}\rangle$ read
\begin{equation}
\begin{split}
\langle E_{i\to o}\rangle =\frac{\displaystyle{\int_{-\frac{\pi}{2}}^{\frac{\pi}{2}}\mathrm{d}\theta} \int_{v_{ib}\left(\theta\right)}^{\infty} \frac{1}{2}mv^2 P_i(v,\theta) \mathrm{d}v }{p_{io}} =\frac{3k_B T_i}{2}+\frac{h r_b \mbox{Erf}\left( \sqrt{\frac{h }{k_B T_i (r_b^2-1)}}\right) e^{-\frac{h}{k_B T_i}}}{p_{io}},
\end{split}
\end{equation}
and
\begin{equation}
\begin{split}
  \langle E_{o\to i}\rangle =\frac{\displaystyle{\int_{-\Theta_{o}}^{\Theta_{o}} \mathrm{d}\theta} \int_{v_{ob}\left(\theta\right)}^{\infty} \frac{1}{2}mv^2P_o(v,\theta) \mathrm{d}v}{p_{oi}} =\frac{3k_B T_o}{2}+\frac{h r_b \mbox{Erf}\left( \sqrt{\frac{h }{k_B T_o (r_b^2-1)}}\right) e^{-\frac{h}{k_B T_o}}}{r_o p_{oi}},
\end{split}
\end{equation}
where $r_b \equiv R_b/R_i$. Moreover,
\begin{equation}
\begin{split}
  p_{io} =\int_{-\frac{\pi}{2}}^{\frac{\pi}{2}}d\theta \int_{v_{ib}\left(\theta\right)}^{\infty} P_i(v,\theta) dv = 1-\mbox{Erf}\left( \sqrt{\frac{h}{k_B T_i}\frac{r_b^2}{r_b^2-1}}\right) +r_b e^{-\frac{h}{k_B T_i}}\mbox{Erf}\left( \sqrt{\frac{h}{k_B T_i}\frac{1}{r_b^2-1}}\right),
  \end{split}
\end{equation}
and
\begin{equation}
\begin{split}
  p_{oi} =\int_{-\Theta_{o}}^{\Theta_{o}} d\theta \int_{v_{ob}\left(\theta\right)}^{\infty} P_o(v,\theta) dv =\frac{1}{r_o}\left( 1-\mbox{Erf}\left( \sqrt{\frac{h}{k_B T_o}\frac{r_b^2}{r_b^2-1}}\right) + r_b e^{-\frac{h}{k_B T_o}} \mbox{Erf}\left( \sqrt{\frac{h}{k_B T_o}\frac{1}{r_b^2-1}}\right) \right),
  \end{split}
\end{equation}

For the averaged time $\langle t_{i\to o}\rangle$ and $\langle t_{o\to i}\rangle$, we have
\begin{equation}
\begin{split}
  \langle t_{i\to o}\rangle &= \frac{p_{ibi}}{p_{io}}t_{ibi} + t_{io}\\
                                    &=\frac{R_i}{p_{io}}\sqrt{\frac{m}{2\pi k_B T_i}}\left(2 r_b^2\Theta_{b}-\pi + 2\sqrt{r_b^2-1} + \int_{-\frac{\pi}{2}}^{\frac{\pi}{2}}\left(L_{io}(\theta)-L_{ibi}(\theta)\right) f_i\left(\theta\right)\mathrm{d}\theta\right)
\end{split}
\end{equation}
and
\begin{equation}
\begin{split}
  \langle t_{o\to i}\rangle &= \frac{p_{oo}}{p_{oi}}t_{oo} + \frac{p_{obo}}{p_{oi}}t_{obo} + t_{oi}\\
                                    &=\frac{1}{p_{oi}}\sqrt{\frac{m}{2\pi k_B T_o}} \left\lbrack \frac{\pi\left( r_o^2-r_b^2\right)}{r_o}R_i +  \int_{-\Theta_{o}}^{\Theta_{o}}L_{oi}(\theta) f_o\left(\theta\right) \mathrm{d}\theta+ 2\int_{\Theta_{o}}^{\Theta_{bo}}L_{oo}(\theta) f_o\left(\theta\right) \mathrm{d}\theta - \int_{-\Theta_{bo}}^{\Theta_{bo}}L_{obo}(\theta) f_o\left(\theta\right) \mathrm{d}\theta\right\rbrack,
\end{split}
\end{equation}
where $\Theta_{b}\equiv \arcsin (R_i/R_b)$, $\Theta_{bo}\equiv \arcsin (R_b/R_o)$, $f_i\left(\theta\right) = e^{-\frac{m v_{ib}^2\left(\theta\right)}{2k_B T_i}}\cos\theta$, and $f_o\left(\theta\right)=e^{-\frac{m v_{ob}^2\left(\theta\right)}{2k_B T_o}}\cos\theta$. To obtain $\langle t_{i\to o}\rangle$ and $\langle t_{o\to i}\rangle$, we need to know $t_{io}$, $p_{ibi}t_{ibi}$, $t_{oi}$, $p_{oo} t_{oo}$, and $p_{obo}t_{obo}$, which are

\begin{equation}
\begin{split}
  t_{io} =\frac{\displaystyle{\int_{-\frac{\pi}{2}}^{\frac{\pi}{2}}\mathrm{d}\theta} \int_{v_{ib}\left(\theta\right)}^{\infty} \frac{L_{io}(\theta)}{v}P_i(v,\theta) \mathrm{d}v }{p_{io}}=\frac{1}{p_{io}}\sqrt{\frac{m}{2\pi k_B T_i}}\displaystyle{\int_{-\frac{\pi}{2}}^{\frac{\pi}{2}}L_{io}(\theta) f_{i}\left(\theta\right) \mathrm{d}\theta},
\end{split}
\end{equation}

\begin{equation}
\begin{split}
  p_{ibi}t_{ibi} &=\int_{-\frac{\pi}{2}}^{\frac{\pi}{2}}\mathrm{d}\theta \int_{0}^{v_{ib}\left(\theta\right)} \frac{L_{ibi}(\theta)}{v}P_i(v,\theta) \mathrm{d}v =\sqrt{\frac{m}{2\pi k_B T_i}}\left(\left(2 r_b^2\Theta_{b}-\pi + 2 \sqrt{r_b^2-1}\right)R_i - \int_{-\frac{\pi}{2}}^{\frac{\pi}{2}}L_{ibi}(\theta)f_{i}\left(\theta\right) \mathrm{d}\theta\right),
\end{split}
\end{equation}
\begin{equation}
\begin{split}
  t_{oi} =\frac{\displaystyle{\int_{-\Theta_{o}}^{\Theta_{o}} \mathrm{d}\theta} \int_{v_{ob}\left(\theta\right)}^{\infty} \left(\frac{L_{oi}(\theta)}{v}P_o(v,\theta)\right) \mathrm{d}v}{p_{oi}} =\frac{1}{p_{oi}}\sqrt{\frac{m}{2\pi k_B T_o}}\int_{-\Theta_{o}}^{\Theta_{o}} L_{oi}(\theta) f_{o}\left(\theta\right) \mathrm{d}\theta,
\end{split}
\end{equation}
\begin{equation}
\begin{split}
  p_{oo} t_{oo} &=\left( \int_{-\frac{\pi}{2}}^{-\Theta_{bo}}  +\int_{\Theta_{bo}}^{\frac{\pi}{2}} \right) \left(\int_{0}^{\infty}  \frac{L_{oo}(\theta)}{v}P_o(v,\theta) \mathrm{d}v\right)\mathrm{d}\theta +\left( \int_{-\Theta_{bo}}^{-\Theta_{o}} +  \int_{\Theta_{o}}^{\Theta_{bo}}  \right) \left(\int_{v_{ob}\left(\theta\right)}^{\infty} \frac{L_{oo}(\theta)}{v}P_o(v,\theta) \mathrm{d}v \right) \mathrm{d}\theta\\
  &=\sqrt{\frac{m}{2\pi k_B T_o}}\left(\frac{\left(\pi-2\Theta_{bo}\right)r_o^2 - 2 r_b \sqrt{r_o^2 - r_b^2}}{r_o}R_i+ 2\int_{\Theta_{o}}^{\Theta_{bo}} L_{oo}(\theta) f_{o}\left(\theta\right) \mathrm{d}\theta \right),
\end{split}
\end{equation}
and
\begin{equation}\label{eq:hne0-tobo}
\begin{split}
  p_{obo}t_{obo} &=\int_{-\Theta_{bo}}^{\Theta_{bo}} \mathrm{d}\theta \int_{0}^{v_{ob}\left(\theta\right)}\left( \frac{L_{obo}(\theta)}{v}P_o(v,\theta)\right) \mathrm{d}v\\
  &=\sqrt{\frac{m}{2\pi k_B T_o}}\left(\frac{2\Theta_{bo}r_o^2+2r_b\sqrt{r_o^2-r_b^2}-\pi r_b^2}{r_o}R_i-\int_{-\Theta_{bo}}^{\Theta_{bo}} L_{obo}(\theta) f_{o}\left(\theta\right) \mathrm{d}\theta\right),
\end{split}
\end{equation}
respectively.
In the limit of $h\to\infty$, the complicated integrals in these expressions can be written down explicitly, so that the current $J^h$ tends to
\begin{equation}
\begin{split}
J^{\infty}&=\lim_{h\to \infty}\frac{\frac{3}{2}Nk_B |T_i - T_o|}{\frac{1}{p_{io}}\sqrt{\frac{m}{2\pi k_B T_i}}\left(2 r_b^2\Theta_{b}-\pi + 2\sqrt{r_b^2-1} \right)R_i  + \frac{1}{r_o p_{oi}}\sqrt{\frac{m}{2\pi k_B T_o}} \left(\pi\left( r_o^2-r_b^2\right)\right)R_i }\\
&= \lim_{h\to \infty}  \frac{\frac{3}{2}N k_B |T_i - T_o|}{\mathrm{Max} \left\lbrace e^{\frac{h}{k_B T_i}}\sqrt{\frac{m}{2\pi k_B T_i}}\left(2 r_b^2\Theta_{b}-\pi + 2\sqrt{r_b^2-1} \right)R_i  , e^{\frac{h}{k_B T_o}}\sqrt{\frac{m}{2\pi k_B T_o}} \left(\pi\left( r_o^2-r_b^2\right)\right)R_i \right\rbrace},
\end{split}
\end{equation}
and as a consequence,
\begin{equation}
\xi^{\infty}=
\frac{\pi\left(r_o^2-r_b^2\right)-\left(2r_b^2 \Theta_{b}-\pi+2\sqrt{r_b^2-1}\right)}{\pi\left(r_o^2-r_b^2\right)+\left(2r_b^2 \Theta_{b}-\pi+2\sqrt{r_b^2-1}\right)}.
\end{equation}

For a fixed $h$, suppose in the limit of $r_o\to \infty$, $\xi^h$ tends to a certain value, denoted as $\xi_{max}^{h}$. Then, we have
\begin{equation}
\xi_{max}^{h} = \frac{\sqrt{T_H} P\left(T_H\right)-\sqrt{T_C} P\left(T_C\right)}{\sqrt{T_H} P\left(T_H\right)+\sqrt{T_C} P\left(T_C\right)},
\end{equation}
where $P(T)= 1 + r_b e^{-\frac{h}{k_B T}} \mbox{Erf}\left( \sqrt{\frac{h}{k_B T}\frac{1}{r_b^2-1}}\right) -\mbox{Erf}\left(r_b \sqrt{\frac{h}{k_B T}\frac{1}{r_b^2-1}}\right)$. It's clear that $\xi_{max}^{h} > \xi_{max}$ for any $h>0$, because
\begin{equation}
\begin{split}
\xi_{max}^{h} &= \frac{\sqrt{T_H} P\left(T_H\right)-\sqrt{T_C} P\left(T_C\right)}{\sqrt{T_H} P\left(T_H\right)+\sqrt{T_C} P\left(T_C \right)}\\
&=\frac{\sqrt{T_H}-\sqrt{T_C}\frac{P\left(T_C\right)}{P\left(T_H\right)}}{\sqrt{T_H}+\sqrt{T_C}\frac{P\left(T_C\right)}{P\left(T_H\right)}}\\
&> \frac{\sqrt{T_H}-\sqrt{T_C}}{\sqrt{T_H}+\sqrt{T_C}}=\xi_{max}.
\end{split}
\end{equation}
The inequality sign is due to the fact that $P(T_C)/P(T_H) <1$, because $\frac{\mathrm{d}P(T)}{ \mathrm{d} T} = \frac{h r_b e^{-\frac{h}{k_B T }} \mbox{Erf}\left(\sqrt{\frac{h}{k_B T \left(r_b^2-1\right) }}\right)}{k_B T^2 } >0$ for any $h>0$.

\section{The 3D counterpart of the generalized Corbino disk with a potential barrier}
Taking the same definitions of $\langle E_{i\to o}\rangle$, $\langle E_{o\to i}\rangle$,
$\langle t_{i\to o}\rangle$, and $\langle t_{o\to i}\rangle$ as in the 2D case,
their expressions are the same as in the latter, given that $\Pi(v, \theta)$ is replaced by $\Pi_\alpha^{3D} (v,\theta)$,
and only the integrals that the integral range of $\theta$ falls in $\lbrack 0, \frac{\pi}{2}\rbrack$ are retained.
We thus have
\begin{equation}
\begin{split}
\langle E_{i\to o}\rangle =\frac{\displaystyle{\int_{0}^{\frac{\pi}{2}}\mathrm{d}\theta \int_{v_{ib}\left(\theta\right)}^{\infty} \frac{1}{2}mv^2 \Pi_i^{3D}(v,\theta) \mathrm{d}v }}{p_{io}} =2k_B T_i+h-\frac{h }{ \left(e^{\frac{h }{k_B T_i\left(r_b^2-1\right) }}-1\right)r_b^2+1},
\end{split}
\end{equation}
\begin{equation}
\begin{split}
  \langle E_{o\to i}\rangle =\frac{\displaystyle{\int_{0}^{\Theta_{o}} \mathrm{d}\theta \int_{v_{ob}\left(\theta\right)}^{\infty} \frac{1}{2}mv^2\Pi_o^{3D}(v,\theta) \mathrm{d}v}}{p_{oi}} =2k_B T_o+h-\frac{h }{\left(e^{\frac{h }{ k_B T_o\left(r_b^2-1\right) }}-1\right)r_b^2+1},
\end{split}
\end{equation}
\begin{equation}
\begin{split}
  \langle t_{i\to o}\rangle &= \frac{p_{ibi}}{p_{io}}t_{ibi} + t_{io}\\
                                    &=\frac{1}{p_{io}}\sqrt {\frac{2\pi m}{ k_B T_i}} \left(\frac{R_i}{6}\left( {r_o^3 - \left(r_o^2-1\right)^{3/2} - 1} \right)+\int_{0}^{\frac{\pi}{2}}f_i^{3D}\left(\theta\right)\left( L_{io}\left( \theta \right)-L_{ibi}\left( \theta \right) \right)\mathrm{d}\theta \right),
\end{split}
\end{equation}
and
\begin{equation}
\begin{split}
  \langle t_{o\to i}\rangle &= \frac{p_{oo}}{p_{oi}}t_{oo} + \frac{p_{obo}}{p_{oi}}t_{obo} + t_{oi}\\
                                    &=\frac{1}{p_{oi}}\sqrt {\frac{2\pi m}{k_BT_o}} \left(\frac{R_i}{6 r_o^2}\left( {r_o^3 + \left(r_o^2-1\right)^{3/2}-1} \right)+\right.\\
                                    ~&\left.\int_{0}^{\Theta_{o}}f_o^{3D}\left(\theta\right)L_{oi}\left(\theta \right)\mathrm{d}\theta +\int_{\Theta_{o}}^{\Theta_{bo}}f_o^{3D}\left(\theta\right)L_{oo}\left(\theta \right)\mathrm{d}\theta - \int_{0}^{\Theta_{bo}}f_o^{3D}\left(\theta\right)L_{obo}\left(\theta \right)\mathrm{d}\theta\right),
\end{split}
\end{equation}
where
\begin{equation}
f_i^{3D}\left(\theta\right)=\left( {\frac{1}{{\sqrt \pi  }}\sqrt{\frac{m v_{ib}^2\left(\theta\right)}{2 k_B T_i}}e^{-\frac{m v_{ib}^2\left(\theta\right)}{2 k_B T_i}} - \frac{1}{2}\mbox{Erf}\left( \sqrt{\frac{m v_{ib}^2\left(\theta\right)}{2 k_B T_i}}\right)} \right)\sin \theta \cos \theta,
\end{equation}
and
\begin{equation}
f_o^{3D}\left(\theta\right)=\left( {\frac{1}{{\sqrt \pi }}\sqrt{\frac{m v_{ob}^2\left(\theta\right)}{2 k_B T_o}}e^{-\frac{m v_{ob}^2\left(\theta\right)}{2 k_B T_o}} - \frac{1}{2}\mbox{Erf}\left( \sqrt{\frac{m v_{ob}^2\left(\theta\right)}{2 k_B T_o}}\right)} \right)\sin \theta \cos \theta.
\end{equation}
To obtain these results, we have substituted $p_{io}$, $t_{io}$, $p_{ibi}t_{ibi}$, $p_{oi}$, $t_{oi}$, $p_{oo} t_{oo}$, and $p_{obo}t_{obo}$, which are, respectively,
\begin{equation}\label{eq:hne0-pio}
\begin{split}
  p_{io} &=\int_{0}^{\frac{\pi}{2}}\mathrm{d}\theta \int_{v_{ib}\left(\theta\right)}^{\infty} \Pi_i^{3D}(v,\theta) \mathrm{d}v = r_b^2 e^{-\frac{h}{k_B T_i}} -\left(r_b^2-1\right) e^{-\frac{h r_b^2}{k_B T_i\left(r_b^2 -1\right)}},
  \end{split}
\end{equation}
\begin{equation}\label{eq:hne0-tio}
\begin{split}
  t_{io} &=\frac{\displaystyle{\int_{0}^{\frac{\pi}{2}}\mathrm{d}\theta} \int_{v_{ib}\left(\theta\right)}^{\infty} \frac{L_{io}(\theta)}{v}P_i(v,\theta) \mathrm{d}v }{p_{io}}\\
  &=\frac{1}{p_{io}}\sqrt {\frac{2\pi m}{{k_B T_i}}} \left(\frac{{{R_i}}}{6}\left( {r_o^3 - \left(r_o^2-1\right)^{3/2} - 1} \right)+\int_{0}^{\frac{\pi}{2}}f_i^{3D}\left(\theta\right){L_{io}}\left( \theta  \right)\mathrm{d}\theta \right),
\end{split}
\end{equation}

\begin{equation}
\begin{split}
  p_{ibi}t_{ibi} &=\int_{0}^{\frac{\pi}{2}}\mathrm{d}\theta \int_{0}^{v_{ib}\left(\theta\right)} \frac{L_{ibi}(\theta)}{v}P_i(v,\theta) \mathrm{d}v \\
  &=-\sqrt{\frac{2 \pi m}{k_B T_i}}\displaystyle{ \int_{0}^{\frac{\pi}{2}}f_i^{3D}\left(\theta\right)L_{ibi}\left(\theta\right) \mathrm{d}\theta},
\end{split}
\end{equation}

\begin{equation}
\begin{split}
  p_{oi} =\int_{0}^{\Theta_{o}} \mathrm{d}\theta \int_{v_{ob}\left(\theta\right)}^{\infty} \Pi_o^{3D}(v,\theta) \mathrm{d}v =\frac{r_b^2 e^{-\frac{h}{k_B T_o}}-\left(r_b^2-1\right) e^{-\frac{h r_b^2}{k_B T_o (r_b^2-1)}}}{r_o^2},
  \end{split}
\end{equation}
\begin{equation}
\begin{split}
  t_{oi} &=\frac{\displaystyle{\int_{0}^{\Theta_{o}} d\theta} \int_{v_{ob}\left(\theta\right)}^{\infty} \left(\frac{L_{oi}(\theta)}{v}\Pi_o^{3D}(v,\theta)\right) \mathrm{d}v}{p_{oi}} \\ &=\frac{1}{p_{oi}}\sqrt {\frac{2\pi m}{k_B T_o}} \left(\frac{R_i}{6 r_o^2}\left( {r_o^3 - \left(r_o^2-1\right)^{3/2}-1} \right)+\int_{0}^{\Theta_{o}}f_o^{3D}\left(\theta\right)L_{oi}\left( \theta  \right)\mathrm{d}\theta \right),
\end{split}
\end{equation}
\begin{equation}
\begin{split}
  p_{oo} t_{oo} &=\int_{\Theta_{bo}}^{\frac{\pi}{2}}\mathrm{d}\theta  \int_{0}^{\infty}  \frac{L_{oo}(\theta)}{v}\Pi_o^{3D}(v,\theta) \mathrm{d}v + \int_{\Theta_{o}}^{\Theta_{bo}} \mathrm{d}\theta \int_{v_{ob}\left(\theta\right)}^{\infty} \frac{L_{oo}(\theta)}{v}\Pi_o^{3D}(v,\theta) \mathrm{d}v \\
  &=\sqrt{\frac{2 \pi m}{k_B T_o}}\left(\frac{ \left(r_o^2-1\right)^{3/2} }{3 r_o^2}R_i + \int_{\Theta_{o}}^{\Theta_{bo}}f_o^{3D}\left(\theta\right)L_{oo}\left( \theta  \right) \mathrm{d}\theta\right),
\end{split}
\end{equation}
and
\begin{equation}
\begin{split}
  p_{obo}t_{obo} &=\int_{0}^{\Theta_{bo}} \mathrm{d}\theta \int_{0}^{v_{ob}\left(\theta\right)}\left( \frac{L_{obo}(\theta)}{v}P_o(v,\theta)\right) \mathrm{d}v\\
  &=-\sqrt{\frac{2 \pi  m}{k_B T_o}}\displaystyle{ \int_{0}^{\Theta_{bo}}f_o^{3D}\left(\theta\right)L_{obo}\left(\theta\right) \mathrm{d}\theta}.
\end{split}
\end{equation}

Again, in the limit $h\to \infty$, $f_i^{3D}\left(\theta\right)=f_o^{3D}\left(\theta\right)=-\frac{1}{2}\sin\theta\cos\theta$, then all the complicated integrals in above expressions can be calculated analytically, so that the heat current tends to
\begin{equation}
\begin{split}
J^{\infty}&=\lim_{h\to \infty}\frac{2Nk_B\left| T_i-T_o\right|}{\frac{1}{p_{io}}\sqrt {\frac{2 \pi m}{k_B T_i}} \left(\frac{R_i}{3}\left(r_b^3 - \left(r_b^2-1\right)^{3/2} - 1 \right)\right)+ \frac{1}{p_{oi}}\sqrt {\frac{2\pi m}{k_B T_o}} \left(\frac{{{R_i}}}{3r_o^2}\left( r_o^3-r_b^3 \right) \right) }\\
&=\lim_{h\to \infty} \frac{2 Nk_B\left| T_i-T_o\right|}{\mathrm{Max}\left\lbrace e^{\frac{h}{k_B T_i}}\sqrt {\frac{2\pi m}{k_B T_i}} \left(\frac{R_i}{3}\left( {r_b^3 - \left(r_b^2-1\right)^{3/2} - 1} \right)\right), e^{\frac{h}{k_B T_o}}\sqrt {\frac{2\pi m}{k_B T_o}} \left(\frac{R_i}{3}\left( r_o^3-r_b^3 \right) \right)\right\rbrace }.
\end{split}
\end{equation}
It leads to
\begin{equation}
\xi^{\infty}= 
\frac{\left( r_o^3-r_b^3 \right) -\left(r_b^3 - \left(r_b^2-1\right)^{3/2} - 1\right)}{\left( r_o^3-r_b^3 \right) +\left(r_b^3 - \left(r_b^2-1\right)^{3/2} - 1\right)}.
\end{equation}

\end{widetext}


\begin{thebibliography}{100}

\bibitem{Peyrard2002}
M. Terraneo, M. Peyrard, and G. Casati,
Phys. Rev. Lett.~\textbf{88}, 094302 (2002).

\bibitem{Peyrard2006}
M. Peyrard,
EPL~\textbf{76}, 49 (2006).

\bibitem{Walker2011}
N. A. Robert and D. G. Walker,
Int. J. Therm. Sci.~\textbf{50}, 648 (2011).

\bibitem{Li2012}
N. Li, J. Ren, L. Wang, G. Zhang, P. H\"anggi, and B. Li,
Rev. Mod. Phys.~\textbf{84}, 1045 (2012).

\bibitem{BCMP2016}
G. Benenti, G. Casati, C. Mej\'{\i}a-Monasterio, and M. Peyrard,
\textit{From thermal rectifiers to thermoelectric devices}, in
\textit{Thermal transport in low dimensions: from statistical physics to nanoscale heat transfer},
Springer Lecture Notes in Physics vol. 921,
edited by S. Lepri (2016).

\bibitem{Wehmeyer2017}
G. Wehmeyer, T. Yabuki, C. Monachon, J. Wu, and C. Dames,
Appl. Phys. Rev.~\textbf{4}, 041304 (2017).

\bibitem{BDLL2023}
G. Benenti, D. Donadio, S. Lepri, and R. Livi,
\textit{Non-Fourier heat transport in nanosystems}, Riv. Nuovo Cim.~\textbf{46}, 105 (2023).


\bibitem{Zhang2010}
L. Zhang, J. Wang, and B. Li,
Phys. Rev. B~\textbf{81}, 100301(R) (2010).

\bibitem{Ouyang2010}
T. Ouyang, Y. Chen, Y. Xie, X. Wei, K. Yang, P. Yang, J. Zhong,
Phys. Rev. B \textbf{82}, 245403 (2010).

\bibitem{Wu2021}
Y. Wu, Y. Yang, L. Lu, T. Wang, L. Xu, Z. Yu, and L. Zhang,
Phys. Rev. E~\textbf{103}, 052135 (2021).

\bibitem{CWZhang2013}
T.-K. Hsiao, H.-K. Chang, s.-C. Liou, M.-W. Chu, S.-C. Lee, and C.-W. Zhang,
Nat. Nanotechnol.~\textbf{8}, 534 (2013).

\bibitem{Volz2019}
R. Anufriev, S. Gluchko, S. Volz, and M. Nomura,
Nanoscale~\textbf{11}, 13407 (2019).

\bibitem{Zardo2020}
D. Vakulov, S. Gireesan, M. Y. Swinkels, R. Chavez, T. Vogelaar, P. Torres, A. Campo, M. De Luca, M. A. Verheijen, S. Koelling, L. Gagliano, J. E. M. Haverkort,
F. X. Alvarez, P. A. Bobbert, I. Zardo, and E. P. A. M. Bakkers,
Nano Lett. \textbf{20}, 2703 (2020).

\bibitem{Song1998}
A. M. Song, A. Lorke, A. Kriele, J. P. Kotthaus, W. Wegscheider and M. Bichler,
Phys. Rev. Lett.~\textbf{80}, 3831 (1998).

\bibitem{Otey2010}
C. R. Otey, W. T. Lau, and S. Fan,
Phys. Rev. Lett.~\textbf{104}, 154301 (2010).

\bibitem{Schmotz2011}
M. Schmotz, J. Maier, E. Scheer, and P. Leiderer,
New J. Phys.~\textbf{13}, 113027 (2011).

\bibitem{Corbino1911}
O. M. Corbino,
Atti R. Accad. Lincei ~\textbf{20}, 342 (1911).

\bibitem{Dolgopolov1992}
V. T. Dolgopolov, A. A. Shashkin, N. B. Zhitenev, S. I. Dorozhkin, and K. von Klitzing,
Phys. Rev. B,~\textbf{46}, 12560 (1992).

\bibitem{Altshuler2020}
A. V. Kavokin, B. L. Altshuler, S. G. Sharapov, P. S. Grigoryev, and A. A. Varlamov,
PNAS,~\textbf{117}, 2846 (2020).

\bibitem{Arrachea2020}
M. Real, D. Gresta, C. Reichl, J. Weis, A. Tonina, P. Giudici, L. Arrachea, W. Wegscheider, and W. Dietsche,
Phys. Rev. Applied \textbf{14}, 034019 (2020).

\bibitem{bath1}
J. L. Lebowitz and H. Spohn,
J. Stat. Phys.~\textbf{19}, 633 (1978).

\bibitem{bath2}
R. Tehver, F. Toigo, J. Koplik, and J. R. Banavar,
Phys. Rev. E~\textbf{57}, R17 (1998).

\bibitem{SM}
The Supplemental Material contains details of the analytical derivation of the heat flux for the (generalized) Corbino disk, as well as
the extension of the heat rectification calculation to the three-dimensional case (motion within a spherical shell) and
to the elliptic Corbino disk.

\end{thebibliography}
\end{document}